# A Perceptual Alphabet for the 10-dimensional Phonetic-prosodic Space

## *a supplement to Oral Billiards*

*Elaine Y L Tsiang*†

Monowave Corporation, Seattle

## Abstract

We define an alphabet, the IHA, of the 10-D phonetic-prosodic space. The dimensions of this space are perceptual observables, rather than articulatory specifications. Speech is defined as a random chain in time of the 4-D phonetic subspace, that is, a symbolic sequence, augmented with diacritics of the remaining 6-D prosodic subspace. The definitions here are based on the model of speech of *oral billiards*, and supersedes an earlier version. This paper only enumerates the IHA in detail as a supplement to the exposition of oral billiards in a separate paper. The IHA has been implemented as the target random variable in a speech recognizer.

## 1. Introduction

It is a fundamental assumption of phonetics that there is a space of articulatory gestures which encompasses all spoken languages. From this space, a particular language selects a set of gestures to represent the bits required to encode its lexicon. This gestural space is universal for all spoken langauges, and is more fine-grained than the acknowledged articulatory repertories of specific languages.

The International Phonetic Alphabet(IPA)[1] defines this gestural space by specifying the gestures at a sufficiently fine resolution to accommodate the production of all spoken languages. We define here the search space for the acoustic emissions from these gestures at a sufficiently fine resolution to cover the perception of all spoken languages. This contrasts with other approaches aimed at constructing acoustic models broad enough to be adaptable to multiple languages[2]

This perceptual space is a product of two subspaces, the phonetic and the prosodic. Our formulation is based on a physical model of articulation as *oral billiards*, rendered sub-maximally observable by acoustic emissions[3][4]. We define the phonetic subspace to be the 4-dimensional value space of the kinematics of oral billiards. We propose an alphabet, the IHear[1] Alphabet(IHA) as the set of symbolic markers in this phonetic subspace. The kinematics of oral billiards give rise to a language-neutral[2] grammar over the IHA. The prosodic subspace consists of 6 directly measurable physical observables which further qualify the acoustic emissions[3]. Together they provide complete coverage of the perceptual space of speech.

†† ter@monowave.com

[1] IHear is a registered trademark of Monowave Corporation

[2] We use the term “language-neutral” rather than “language-independent” to mean independence from the characteristics of any particular language, but not from what we understand to be a spoken language in general.

## 2. The phonetic alphabet

The IHA phonetic alphabet results from the conjugation of a *phthongal alphabet* defined in [3], with the 8 manners of *closure*, *nasal*, *plosive*, *fricative*, *approximant*, *semivowel*, *vowel* and *glottalPlosive*. The phthongal alphabet is characterized by the phonetic attributes of *place*, *frontBack and openClose*, and is based on a broken symmetry that allows each of the three values of *place* to take on a 2D subspace of *frontBack* ⊗ *openclose*. Below we enumerate in complete this conjugation.

We also contrast the IHA with the more recent but unofficial IPA[5], pointing out the similarities and differences.

### 2.1. The phonetic subspace

The manners n*asal*, *fricative*, *approximant* and *vowel* correspond with the IPA manners of the same name. *GlottalPlosive*, referred to as the glottal stop, and the epiglottal version are included under plosives in the IPA. In the IHA, a glottalPlosive is defined as a separate manner. It has the same relation of a fast onset to a vowel as a non-glottal plosive has to a fricative. The IHA does not have values for the IPA manners of taps, trills and the two lateral manners[3]. Taps have been relegated to their corresponding approximants. The time resolution of the phonetic markers will be fine enough to mark the trills as sequences of alternating closures and plosives. The phones in the lateral manners of the IPA will be discussed in the sections detailing the proposed markers. The manner of *semivowel* has the same relation to *vowel* as *approximant* has to *fricative*.

There are 5 values for frontBack: *front*, *frontLike*, *central*, *backLike and back*, same as the IPA, but with slightly different naming. Similarly, openClose has 7 values: *close*, *closeLike*, *closeMid*, *mid*, *openMid*, *openLike and open*.

The set of values for place is where the IHA differs most from the IPA. There are 3 values: *palatalAlveoLabial(PAL)*, *velarUvualrPharyngeal(VUP) and glottal(GLOT)*. Each place has its own 3-dimensional subspace of manner⊗frontBack⊗ openClose, as will be detailed in the following sections. The 8 places of articulation from bilabial to palatal in the IPA have been absorbed into the one value, PAL, of place, and assigned frontBack and openClose values. Similarly, the 4 places from velar to epiglottal have also acquired frontBack and openClose values. The broken symmetry between PAL and VUP reflects the broken symmetry of the oral cavity.

[3] Nor does the IHA currently deal with the non-pulmonic phones due to lack of data. In terms of oral billards, this category will likely eventually fall under the prosodic dimension of voicing.

## 2.2. The phonetic markers

Table 1 through Table 11 exhibit charts for these subspaces. We do not exhaustively show all markers. Where the markers are represented by similar subscripting or superscripting, we simply declare them without explicit tabulation. For example, the closures for all values of place have the same frontBack⊗openClose subspaces as the corresponding vowels or fricatives, and are represented by subscripting the corresponding symbol with “$_{o}$”.

Note that we do not distinguish between voiced and unvoiced phones, and the symbols that have voicing connotation in the IPA do not have such connotation in the IHA. It is well-known that the voiced vs. unvoiced distinction in different languages has varying voice onset times[6]. Also identifying voicing with periodicity is problematic. Whispered speech, for example, retains the perceptual difference between voiced and unvoiced phones. Instead, we define voicing as the presence or absence of the first formant. The onset and offset times and the degree of presence of the first formant will be a diacritic. From this information, and possibly other prosodic markings, it is up to the specific language to define what it considers voiced, or unvoiced, or the number of voicing categories. This implies that all IHA phones may have voiced and unvoiced versions, including nasals, approximants and even vowels. In addition, closures constitute a full manner, with a full complement of distinct phones as any other manner.

The symbols are usually of the unvoiced variety in the IPA, except for approximants, where some symbols are recommissioned from taps or the voiced fricatives. We use vowel superscripts to indicate frontBack and openClose, except for the PALs. In general, we aim for a single unicode point, instead of combinations.

### 2.2.1. *The vowels*

The IHA regards the vowels as glottal, by virtue of which the traditional great divide between vowels and consonants becomes a difference in place[4]. However, the vowels *are* unique in that they have no other place than glottal, and only the glottal place has vowel entries.

Table 1. *Vowels.*

| | front | frontLike | central | backLike | back |
|---|---|---|---|---|---|
| **close** | i | | ɨ | | u |
| **closeLike** | | ʏ | ɪ̈ | ʊ | |
| **closeMid** | e | | ɘ | | o |
| **mid** | | | ə | | |
| **openMid** | | ɛ | ɜ | | ɔ |
| **openLike** | | æ | ɐ | | |
| **open** | | a | | | ɑ |

### 2.2.2. *The GLOTs*

The glottal vowel subspace *frontBack⊗openClose* is identical with the IPA vowel space. There is a one-to-one correspondence between the IPA vowels and the IHA vowels. However, rounding is not a binary-valued variable (as in traditional “fronting” or “rounding”) in the IHA. Instead, rounding, like voicing, is a prosodic dimension, and takes on a range of values. See section 3 below.

The vowels are, however, less unique in the IHA than the IPA. First, the glottal frontBack⊗openClose subspace as shown in Table 1 also endows the other manners of closure, glottalPlosive and semivowel, yielding almost as many closures, glottalPlosives and semivowels as vowels. The IPA diacritic markings of glottalization is simply a separate manner with the same range of frontBack and openClose values as vowels. Symbols for glottal closures are obtained by subscripting the corresponding vowel symbol with “$_{o}$”, for example, “a$_{o}$”; glottalPlosives by superscripting “ʔ” with the corresponding vowel, as shown in Table 2; and semivowels by similarly superscripting “ɦ”, as shown in Table 3. The semivowel subspace is seen to be reduced from that of vowel, missing some extremal markers, “i”, “u” and “ɑ”, the so-called “quantal” vowels, due to the status of a semivowel being a non-continuant and an approximation to a vowel.

We note that the IPA glottal fricative is absent from the IHA. Its apparent disappearance is due to the relegation of voicing to a prosodic status[3]. The glottal fricative, usually realized as a completely or partially de-voiced vowel, is simply a vowel in the IHA. This means that the glottal fricative, Instead of disappearing, has in effect acquired as many different versions as the vowel.

Although there are no glottal nasals in the IPA or the IHA, glottal nasals may be regarded as nasalized vowels, and like the glottal fricatives, a glottal nasal is realized as a prosodic diacritic of vowels.

Table 2. *GlottalPlosives.*

| *fricatives* | front | frontLike | central | backLike | back |
|---|---|---|---|---|---|
| **close** | $ʔ^{i}$ | | $ʔ^{ɨ}$ | | $ʔ^{u}$ |
| **closeLike** | | $ʔ^{ʏ}$ | $ʔ^{ɪ̈}$ | $ʔ^{ʊ}$ | |
| **closeMid** | $ʔ^{e}$ | | $ʔ^{ɘ}$ | | $ʔ^{o}$ |
| **mid** | | | $ʔ^{ə}$ | | |
| **openMid** | | $ʔ^{ɛ}$ | $ʔ^{ɜ}$ | | $ʔ^{ɔ}$ |
| **openLike** | | $ʔ^{æ}$ | $ʔ^{ɐ}$ | | |
| **open** | | $ʔ^{a}$ | | | $ʔ^{ɑ}$ |

Table 3. *Semivowels.*

| *fricatives* | front | frontLike | central | backLike | back |
|---|---|---|---|---|---|
| **close** | | | | | |
| **closeLike** | | $ɦ^{ʏ}$ | $ɦ^{ɪ̈}$ | $ɦ^{ʊ}$ | |
| **closeMid** | $ɦ^{e}$ | | $ɦ^{ɘ}$ | | $ɦ^{o}$ |
| **mid** | | | $ɦ^{ə}$ | | |
| **openMid** | | $ɦ^{ɛ}$ | $ɦ^{ɜ}$ | | $ɦ^{ɔ}$ |
| **openLike** | | $ɦ^{æ}$ | $ɦ^{ɐ}$ | | |
| **open** | | | | | |

### 2.2.3. *The VUPs*

The VUP frontBack⊗openClose subspaces are more constricted than the GLOT ones. We show the VUP fricatives, plosives, nasals and approximants in Table 4 through Table 7. The labialized velar, which is regarded as co-articulated and presented as approximants outside of the IPA chart for pulmonic consonants, is simply the close back velar in the IHA. The IPA lateral velar is likewise the closeLike backLike velar (thus reassigning one of the phones in the IPA lateral manners). Because of the superscript notation, we have included the corresponding IPA symbols in parentheses. The velars with different frontBack and openClose values constitute the IPA velarization diacritic. On the other hand, the front and frontLike velar fricatives and approximants also serve as the IPA diacritic palatalization of velars. The uvulars frequently occur as “lazy” or fast articulations of velars in languages that do not use the velar/uvular contrast.

[4]See Oral Billiards[3] for the explanation of the site of glottal being at the jaw line, rather than at the glottis.

Table 4. VUP *fricatives*.

| *fricatives* | front | frontLike | central | backLike | back |
|---|---|---|---|---|---|
| close | xⁱ | | | | xᵘ(ʍ) |
| closeLike | | xᶦ | xᵊ | xᵒ(ʟ) | |
| closeMid | | χᶦ | χᵊ | χᵒ | |
| mid | | | ħᵊ | ħᵒ | |

Table 5. VUP *plosives*.

| *plosives* | front | frontLike | central | backLike | back |
|---|---|---|---|---|---|
| close | kⁱ | | | | kᵘ |
| closeLike | | kᶦ | kᵊ | kᵒ | |
| closeMid | | qᶦ | qᵊ | qᵒ | |
| mid | | | ʡᵊ | ʡᵒ | |

Table 6. VUP *nasals*.

| *nasals* | front | frontLike | central | backLike | back |
|---|---|---|---|---|---|
| close | | | | | |
| closeLike | | ŋᶦ | ŋᵊ | ŋᵒ | |
| closeMid | | ɴᶦ | ɴᵊ | ɴᵒ | |

Table 7. VUP *approximants*.

| *approximants* | front | frontLike | central | backLike | back |
|---|---|---|---|---|---|
| close | | | | | |
| closeLike | | ɰᶦ | ɰᵊ | ɰᵒ(w) | |
| closeMid | | ʁᶦ | ʁᵊ | ʁᵒ(ʟ) | |
| mid | | | ʕᵊ | ʕᵒ | |

### *2.2.4. The PALs*

The palatal, alveolo-palatal, and the palato-alveolar IPA phones are assigned the value of front or frontLike. Likewise, the labials are assigned the back or backLike values. The central PALs, assigned increasing values of openClose, correspond to the dental, alveolar and retroflex IPA phones. The symbol [ɻ], used to represent the IPA retroflex approximant, has been recommissioned to represent the marker for a central mid fricative PAL, more familiar to American English listeners as the sound of the grapheme <r>, but represented in its approximant version in both the IHA and the IPA by the inverted [ɹ]. The lateral alveolar, [ɬ], designated as belonging to a separate lateral manner in the IPA, are assigned the values mid backLike, due to its relation to the other PALs being analogous to that of the mid backLike [ʕᵒ] to the other VUPs. The IPA lateral retroflex sits back of the retroflex, and assigned the values closeMid backLike.

In the IHA, the IPA affricates have become the plosive versions of the corresponding fricatives that have no direct plosives in the IPA. Affricates, being homorganic, are perceptually indistinguishable from sounds produced by excising the initial portion of their fricative versions (or final portion for the time-reversed cases). Likewise, the other PAL plosives, the bilabial, dental, retroflex and palatal plosives, can also be produced by a fast onset/offset to their fricative versions. For certain phones, such as [s], there is no independent plosive-only realization. For such phones, the affricates stand in place of plosives. This generalizes to other PALs such as [ɻ] and the laterals as well. We use a combining caron for these less common plosives.

Table 8. *PAL plosives*.

| *plosives* | front | frontLike | central | backLike | back |
|---|---|---|---|---|---|
| close | c | tɕ | t | | p |
| closeLike | | tʃ | ts | ʔ | |
| closeMid | | ʎ̆ | ʈ | ɭ̆ | |
| mid | | | ɻ̆ | ɬ̆ | |

Table 9. *PAL fricatives*.

| *fricatives* | front | frontLike | central | backLike | back |
|---|---|---|---|---|---|
| close | ç | ɕ | θ | | ɸ |
| closeLike | | ʃ | s | f | |
| closeMid | | ʎ | ʂ | ɭ | |
| mid | | | ɻ | ɬ | |

Table 10. *PAL nasals*.

| *nasals* | front | frontLike | central | backLike | back |
|---|---|---|---|---|---|
| close | | | | | m |
| closeLike | | ɲ | n | ɱ | |
| CloseMid | | | ɳ | | |

Table 11. *PAL approximants*.

| *approximants* | front | frontLike | central | backLike | back |
|---|---|---|---|---|---|
| close | | | | | |
| closeLike | | j | ɾ | ʋ | |
| closeMid | | ɿ | ɽ | ɺ | |
| mid | | | ɹ | l | |

## 2.3. Co-articulation

The affricates have been placed into the plosive subspaces. The labialized velars are already in the velar subspace. Other so-called co-articulations will be strictly ordered in the time-aligned output from the IHA phone recognizer. The times of two consecutive phones may be very close, but no two phones will be emitted at the same time.

## 2.4. Some phonetic diacritics

The IHA does not have the full set of phonetic diacritics in the IPA. Only glottalization, pharyngealization, velarization and some palatalization are already marked by virtue of the frontBack⊗openClose subspaces as discussed above. Nasalization, however, will be marked, like voicing, with the onset and offset times of the nasal formants and a variable representing the degree of their presence.

We will discuss rounding in more detail below.

# 3. The prosodic subspace

By virtue of the presentation of a time-aligned random chain of IHA phones, durations(D) are directly computable from the time alignment. Instead of defining stress, a complex observable, we present the loudness(L) of the phone. Tone (T) is simply pitch. D, T and L will all be in quantized logarithmic units. Voicing(V) will be a diacritic whose value represents the degree of presence of the first formant. Similarly, nasalization(N) represents the degree of presence of nasal formants. Rounding (R) is defined to be the change over time in the logarithm of the effective vocal tract length, which is usually assumed to be a constant per speaker, but is in fact dynamic down to the inter-syllable scale. Rounding is thus given in the quantized logarithmic unit of a semitone. A positive value indicates “fronting” or “un-rounding”, and a negative value, “rounding”. They constitute the 6 dimensions of the prosodic subspace.

Our division between the phonetic and prosodic does not coincide exactly with their conventional distinction. Nor do duration, loudness and pitch coincide with the conventional meanings of syllable length, stress and intonation. Voicing is conventionally not regarded as prosodic, and nasalization and rounding are regarded as possibly supra-segmental, but not prosodic. However, none of the conventional notions of syllable length, stress, intonation, voicing or rounding are computationally well defined. The merit of the 6 prosodic dimensions is that they are all computationally well-defined.

The value space for an IHA phone is then the 10-dimensional $manner \otimes frontBack \otimes openClose \otimes place \otimes R \otimes N \otimes V \otimes T \otimes D \otimes L$. We complete this space with one point of origin, the null phone.

## 4. Conclusions

We have formulated a 10-dimensional phonetic- prosodic space and dfined an alphabet, the IHA, as a set of markers in the phonetic subspace of $place \otimes frontBack \otimes openClose \otimes manner$ . The alphabet, with the language-neutral grammar of the syllable[3], and the diacritics over the phone sequences, can be used to systematically construct the phnology, and thereby higher liguistic levels, of any specific spoken language.